\documentclass[journal]{IEEEtran}
\usepackage{amsmath,amssymb,amsfonts}
\usepackage{algorithmic}
\usepackage{array}
\usepackage[pdftex]{graphicx}
\graphicspath{ {images/} }
\usepackage[caption=false,font=footnotesize]{subfig}
\usepackage{textcomp}
\usepackage[table,xcdraw]{xcolor}
\usepackage[T1]{fontenc}
\usepackage{enumitem}
\usepackage{cite}
\usepackage{booktabs}
\usepackage{comment}

\begin{document}

\title{Natural Language Communication with a Conversational Teachable Agent}

\author{Rachel Love$^{1}$, Edith Law$^{2}$, Philip R. Cohen$^{3}$, and Dana Kuli{\'c}$^{4}$ 

\thanks{$^{1}$Rachel Love is with the Department of Electrical and Computer Systems Engineering, Monash University, Australia
    {\tt\footnotesize rachel.love@monash.edu}}%
\thanks{$^{2}$Edith Law is with the David R. Cheriton School of Computer Science, University of Waterloo, Waterloo, Canada
    {\tt\footnotesize edith.law@uwaterloo.ca}}%
\thanks{$^{3}$Philip R. Cohen is with Openstream, Inc., and the Department of Data Science and Artificial Intelligence, Faculty of Information Technology, Monash University, Australia
    {\tt\footnotesize philip.cohen@monash.edu}}%
\thanks{$^{4}$Dana Kuli{\'c} is with the Department of Electrical and Computer Systems Engineering and Department of Mechanical and Aerospace Engineering, Monash University, Australia
    {\tt\footnotesize dana.kulic@monash.edu}}%
}

\maketitle

\begin{abstract}
Conversational teachable agents offer a promising platform to support learning, both in the classroom and in remote settings. In this context, the agent takes the role of the novice, while the student takes on the role of teacher. This framing is significant for its ability to elicit the Prot\'eg\'e effect in the student-teacher, a pedagogical phenomenon known to increase engagement in the teaching task, and also improve cognitive outcomes. In prior work, teachable agents often take a passive role in the learning interaction, and there are few studies in which the agent and student engage in natural language dialogue during the teaching task. This work investigates the effect of teaching modality when interacting with a virtual agent, via the web-based teaching platform, the Curiosity Notebook. A method of teaching the agent by selecting sentences from source material is compared to a method paraphrasing the source material and typing text input to teach. A user study has been conducted to measure the effect teaching modality on the learning outcomes and engagement of the participants. The results indicate that teaching via paraphrasing and text input has a positive effect on learning outcomes for the material covered, and also on aspects of affective engagement. Furthermore, increased paraphrasing effort, as measured by the similarity between the source material and the material the teacher conveyed to the robot, improves learning outcomes for participants.

\end{abstract}


%
\IEEEpeerreviewmaketitle

\section{Introduction}
\IEEEPARstart{A}{ugmenting} classroom teaching with technology can support students with personalised educational needs, while mitigating limited teaching resources. Educational pedagogical agents aim to deliver personalised learning interactions with students. Additionally, pedagogical agents have the flexibility to be deployed in any setting, which is beneficial given the demands and challenges of remote learning exposed in recent times.

\begin{figure}[!t]
    \centering
    \includegraphics[width=\linewidth]{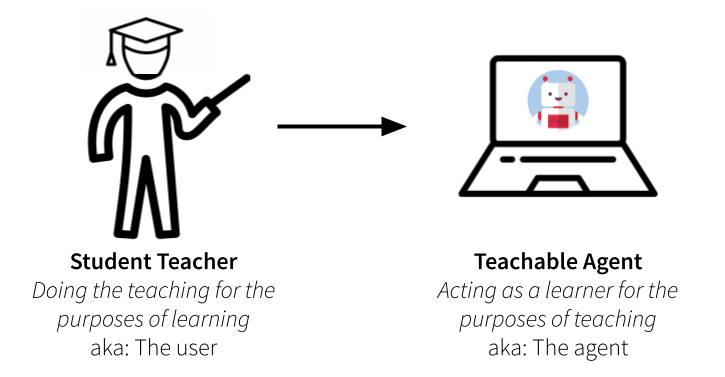}
    \caption{Illustration to show the roles in an interaction with a teachable agent. The user takes on the roles of teaching for their own learning. Images via Vecteezy (https://www.vecteezy.com/)}
    \label{teachable_agent_roles}
\end{figure}

Within this area, both virtual and embodied agents have been deployed to engage students on a range of subjects, and occupying roles as both the tutor and the novice \cite{belpaeme_social_2018}. When pedagogical agents act in the role of a novice, or learner, the student takes on the role of a teacher, communicating material to the agent, for the benefit of the student's own learning. An illustration of the roles in this interaction is given in Figure \ref{teachable_agent_roles}. The use of this framing aims to elicit the Prot\'eg\'e Effect \cite{chase_teachable_2009}, a pedagogical phenomenon in which the student is likely to invest more in learning the material when it is for the benefit of the agent. Students synthesise information better, and adapt their teaching based on the tutee's performance, which may lead to positive cognitive outcomes \cite{biswas_learning_2005}. Allowing the agent to fill the role of pupil may also act in the agent's own favour, as its actions may not always be perfect, and any errors may be more easily forgiven if it is not expected to be an expert.

Teachable agents rarely engage in conversations with their student teachers, while intelligent tutoring systems employ this style of interaction more frequently in order to emulate tutoring dialogues \cite{graesser_autotutor_1999,ward_my_2013}. When natural language is not the primary mode of communication, teachable agents may instead use buttons \cite{yadollahi_when_2018}, or a concept map \cite{biswas_learning_2005}. When learning is considered along the continuum of active (doing something), constructive (producing something) to interactive (exchanging information with someone), students have been shown to experience greater learning gains when interacting with material through discourse with a peer or tutor \cite{ward_my_2013}. Allowing students to interact with a teachable agent in this way may provide the same benefits to learning. Using natural language to synthesise and communicate new information also draws on the benefits of paraphrasing as a comprehension strategy \cite{kletzien_paraphrasing_2009}. In pedagogical research, paraphrasing has been shown to encourage students to connect new material with prior knowledge, and establish retrieval cues \cite{kintsch_comprehension_1998}.

The Curiosity Notebook \cite{lee_curiosity_2021} is an example of a technology that supports conversational teaching interactions with a teachable agent, either virtually, or via integration with a robot. The Curiosity Notebook is a learning by teaching web application, in which students teach an agent about a classification task. In the current implementation, there is limited communication through natural language to direct the teaching interaction. Users can communicate to the teachable agent by clicking on one of seven teaching buttons, after which they are prompted to teach material through a series of agent-directed questions. To these questions, users sometimes type short, free-form answers, but largely are prompted to select sentences from the source material.

This study builds upon the prior work conducted with the Curiosity Notebook. In this work, we first investigate the effect of teaching modality on learning outcomes and engagement, during an interaction with a virtual teachable agent.  In particular, the study investigates the effects of {\it rephrasing} by comparing two teaching modalities: (1) by selecting full sentences from source material in order to teach the agent information \cite{lee_curiosity_2021}, against (2) the student typing their teaching utterances into a chat window, with encouragement to put the source material into their own words. Our results show that teaching modality influences learning outcomes and engagement, and that the amount of rephrasing effort correlates to learning gains. 

\section{Related Work}

There has been significant development of educational technologies that aim to address the need for widely accessible, personalised learning tools for the classroom. Pedagogical agents are one such technology: virtual or embodied characters designed to help students learn material. Such agents can occupy a variety of roles, for example a tutor \cite{graesser_autotutor_1999, moreno_case_2001}, a peer learner \cite{roselyn_lee_case_2007}, or a novice \cite{biswas_learning_2005,brophy_teachable_1999,matsuda_cognitive_2012}.  The following sections provide an overview of educational teachable agents, research relevant to teaching via natural language communication, an introduction to the Curiosity Notebook \cite{lee_curiosity_2021}, the technology upon which this research is based, and an overview of the measures related to this study.

\subsection{Educational Teachable Agents}
Much of the initial research in the field of educational technology was focused on the development of Intelligent Tutoring Systems (ITS) \cite{wenger_artificial_1987}. These systems were designed to provide instruction similar to that of a knowledgeable human tutor \cite{ward_my_2013}. Early approaches were limited by not including the student as an active participant in the learning interaction \cite{wenger_artificial_1987}, though later work has remedied this. The platform AutoTutor \cite{graesser_autotutor_1999} supports conversations between a virtual tutor and student, where the virtual tutor is capable of engaging in human-inspired tutoring dialogue. The platform has been shown to produce learning gains across a variety of domains \cite{nye_autotutor_2014}.

More recently, pedagogical agents have been developed which occupy different roles than those in ITS, such as where the agent acts as a learner, or novice. In such interactions, the student takes on the role of a teacher, engaging in the learning-by-teaching paradigm, a recognised pedagogical tool \cite{werfel_embodied_2014}. Learning by teaching is beneficial to students as it elicits the Protégé Effect, where a student is likely to invest more in learning the material when it is for the benefit of someone else \cite{chase_teachable_2009}. Studies have shown that teaching someone else promotes more organised cognitive structures, compared to learning for oneself \cite{bargh_cognitive_1980}. Additionally, it has been shown that teaching requires the student to reflect upon their teaching based on the performance of the tutee, which may lead to positive cognitive outcomes \cite{biswas_learning_2005}. The presence of the Prot\'eg\'e effect has been confirmed in several studies, with positive cognitive \cite{nestojko_expecting_2014} and meta-cognitive \cite{muis_learning_2015} outcomes.

Applications of the learning-by-teaching approach include SimStudent \cite{matsuda_cognitive_2012}, a tool to help students gain skills in mathematical problem solving. In a game-like interaction, a virtual agent performs steps to solve a mathematical problem, and the student provides hints and corrections to help the agent arrive at the correct answer. Another example is Betty's Brain \cite{biswas_learning_2005}, in which students teach Betty, a virtual learner, about casual relationships in science through the manipulation of concept maps in a shared visual interface. This approach has been shown to improve motivation and learning gains.

These teachable agents share a limited use of natural language as the mode of teaching, relying largely on interaction with the computer interface. Research shows that when human-computer interfaces are consistent with social conventions in daily life, this leads to a more engaging and satisfying user experience \cite{ward_my_2013}. This motivates the use of tutoring dialogues in technology such as AutoTutor, as this is the mode of interaction between human tutors and students \cite{graesser_autotutor_1999,nye_autotutor_2014}. There are also benefits to learning outcomes when students engage with material through discourse and argumentation with a peer or tutor \cite{ward_my_2013}. Technologies such as AutoTutor support complex dialogue for a virtual tutor, however there is a lack of such dialogue systems for teachable agents.

\subsection{Teaching via Natural Language}
Interacting with a teachable agent using natural language not only elicits the Prot\'eg\'e effect, and its attendant benefits, but also engages the student-teacher in paraphrasing, which is recognised as an effective comprehension strategy in the pedagogical research \cite{kletzien_paraphrasing_2009}. The act of reading and paraphrasing content helps to establish retrieval cues, and encourages the student to connect more deeply to the material \cite{kintsch_comprehension_1998}. 

Natural Language Processing (NLP) is a necessary component for a teachable agent taught via natural language conversation. NLP uses computational techniques in order to learn, understand, and produce human language content \cite{hirschberg_advances_2015, torfi_natural_2021}. A review of NLP approaches \cite{cambria_jumping_2014} groups them into techniques concerned with syntax, semantics, and pragmatics, which respectively are concerned with grammar, word meaning, and word meaning with context. Dialog systems can use statistical NLP approaches like intent classification and slot filling, for example Snips \cite{coucke_snips_2018}, which uses logistic regression to train intent classifiers, and several linear-chain conditional random fields for slot extraction. Alternatively, approaches based on lexical semantics such as Latent Semantic Analysis can be used, and can be seen in dialog systems such as the previously mentioned AutoTutor \cite{masche_review_2018,graesser_autotutor_1999}.

\subsection{Curiosity Notebook}
The Curiosity Notebook \cite{law_curiosity_2020} is one such technology that has the potential to support an interaction with a teachable robot, through natural language, and it forms the basis of this research. The Curiosity Notebook is a highly configurable, learning by teaching web application, where students teach an agent about different classification tasks. It supports flexible agent embodiment---for example a virtual agent \cite{chhibber_towards_2019}, or a physical robot \cite{chhibber_using_2019}---the configuration of agent characteristics---for example its style of humour \cite{ceha_can_2021}---and also facilitates group based teaching \cite{ravari_effects_2021}. In the current implementation, natural language input is only supported for short, defined inputs, such as what topic to teach or responses to well-defined questions that are easy to parse (e.g. what kind of rock is Granite?). Teaching the agent about the features of these items occurs through the selection of sentences within source articles embedded in the interface. The research described in this paper extends upon this implementation by adding support for natural language communication throughout the teaching interaction.

\subsection{Student Engagement and Learning Outcomes}
Education technologies are typically evaluated along two main metrics: the efficacy of the tool in teaching students, and the students' interaction with the system \cite{jenkinson_measuring_2009}. These metrics consider, respectively, the knowledge gain or learning outcomes, and the functionality of the technology (independently of learning outcomes). A review of educational technologies found that almost 78.6\% of included studies measured learning \cite{lai_how_2019}. Additionally, almost 61.6\% measured affective elements, such as perceptions, engagement and attitudes and beliefs \cite{lai_how_2019}. Educators are particularly concerned with educational technologies increasing student engagement \cite{bond_mapping_2020}. Fredricks et al. \cite{fredricks_school_2004} conceptualise student engagement in three dimensions: behavioural, which considers observable behaviours related directly to the learning process, affective, the emotional response to the educational experience, and cognitive, the expenditure of energy related to comprehension and learning.

\subsection{Summary}
Teachable agents have been shown to improve learning outcomes over tutoring agents, through the Prot\'eg\'e effect \cite{chase_teachable_2009,biswas_learning_2005,nestojko_expecting_2014}. They have been used in applications where students manipulate interface elements such as concept maps \cite{biswas_learning_2005}, or provide hints to a virtual student \cite{matsuda_cognitive_2012}, rarely using natural language as a teaching modality. Intelligent tutoring systems such as AutoTutor \cite{graesser_autotutor_1999} have successfully employed natural language in tutoring conversations, though natural language dialogue systems for teachable agents remain an under researched area.

The Curiosity Notebook \cite{law_curiosity_2020} supports teaching conversations between a student and teachable agent, however its natural language capabilities to date are limited to very short inputs. This work builds upon this technology to examine the effects of teaching modality on learning outcomes and engagement.

\section{Methodology}
The aim of this study was to investigate the effect of teaching modality within an interaction with a virtual teachable agent. The study compared two methods of providing information to the teachable agent: (1) by selecting full sentences from source material in order to teach the agent information, as implemented in the original implementation of the Curiosity Notebook \cite{law_curiosity_2020}, against (2) the student paraphrasing the source material by typing their teaching utterances into a chat window, with encouragement to put the source material into their own words. In the first condition, the agent was named \emph{Alpha}, and the condition will be described as the sentence selection condition; in the second condition the agent was named \emph{Gamma}, and the condition will be described as the text input condition.

The study aimed to answer how the interaction modality affects two main metrics: Firstly, the student-teacher's learning outcomes, and secondly, their engagement in the teaching task. As discussed in the related works, student engagement can be considered along three dimensions: behavioural, affective, cognitive. It is expected that the intervention of teaching modality would most significantly effect aspects of behavioural engagement, and also the participants' perception of the teaching task.

The implementation required the development or revision of the following components: the Curiosity Notebook interface to support natural language input, a natural language model to parse user input utterances, and agent utterance generation in order to drive the interaction and reflect the different teaching modalities.

\subsection{Curiosity Notebook Interface}

\begin{figure*}[!t]
\centerline{\includegraphics[width=\linewidth]{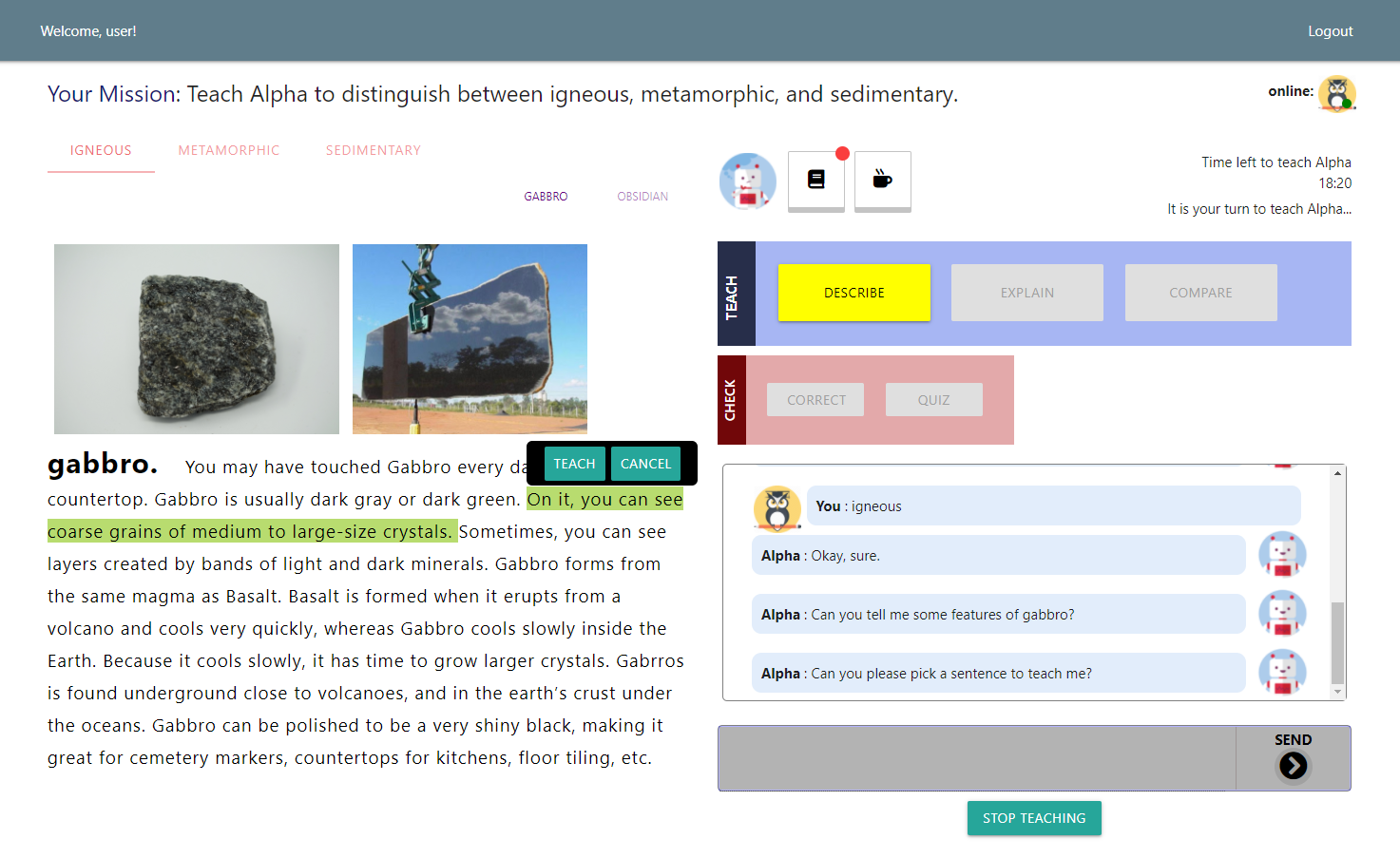}}
\caption{The Curiosity Notebook interface, in the sentence selection condition. The user has clicked on the highlighted sentence, and it is available to teach}
\label{curiosity_notebook}
\end{figure*}

The Curiosity Notebook is a learning-by-teaching platform built as an Angular web application, with Python Flask on the backend \cite{law_curiosity_2020}. The teaching activity is a classification task, in this experiment on rocks and minerals.  In this study, we used a simplified version of the Curiosity Notebook; the interface can be seen in figure \ref{curiosity_notebook}. There are three rock categories, with two examples within each category, for a total of six rocks for each test condition, and twelve in total. Each rock is accompanied by a short article, written in simple language, and a picture. Contained within the article is information relating to a total of 30 rock features. 

Three teaching actions are available: `Describe’, in which the user teaches a feature of the rock, `Explain’, in which the user explains why a feature is present in the rock, and `Compare’, in which the user compares features that are similar, or different, between two rocks. Users can quiz the agent on the content it has learned, and correct information it has mis-learned. A button can be clicked to show the agent’s ``notebook’’ which shows what the agent has learned.

\subsection{Teaching via Sentence Selection}
While teaching Alpha in the sentence selection condition, users communicate in one of two ways. The user will type in the chatbox using natural language when prompted for simple information such as which rock they would like to teach. The information in these utterances is recognised using a simple string match for known entities e.g. recognising `igneous' in the input `this is an igneous rock.'

When the user is required to select a sentence to teach the agent new information, the chatbox is greyed out, and when the student hovers their mouse over the sentences in the source material, they animate with a yellow highlight effect, and are clickable in order to communicate the sentence to the agent.

If a sentence is about the feature of an object (e.g., colour of Slate), that sentence will be mapped to a set of feature IDs.  When a sentence is selected, the features are added to the agent’s database, and each appear as a `note’ in the agent's notebook, which keeps track of what it has learned. In order to avoid Alpha being a perfect learner, and therefore seem cleverer than Gamma, an artificial learning error was included in the sentence selection condition such that 20\% of taught sentences resulted in an error response that prompted the student to select something different.

This version of the interface was used as the baseline condition in the experiment.

\subsection{Teaching via Text Input}
While teaching Gamma, all communication occurs via the user typing in the chatbox, and sentences in the articles are not clickable at any time.

User inputs are parsed using a natural language model built using the Snips Natural Language Understanding python library \cite{coucke_snips_2018}. The library supports parsing input sentences written in natural language and extracting structured information such as intents and slot values.

The natural language model used in the text input condition was designed in such a way that input utterances would be mapped onto the 30 known features already in the database. A combination of intents and slot values was created to match inputs to the known database features. These were named in alignment with the feature ID they corresponded to i.e. the slot value for `large crystals’’ was named ``2’’ to match its feature ID, while the intent ``why the rock has large crystals’’ was named ``12’’ to match its feature ID.

For each input sentence, Snips extracts all matched intents, and if the intent has slots defined, the slot values as well, ordered by confidence in the match. The array of matched intents was looped over and those with a confidence score of 20\% or greater were included as features to add to Gamma’s notebook. 

The language model was developed by hand, using the source material as a guide to generate micro and macro variations of the content within them. Input utterances obtained during the pilot phase of this experiment were used to supplement the model, which then remained the same for all participants in this experiment.

\subsection{Mixed-Initiative Interaction}
A mixed-initiative capability was added in this experiment. In the published version of the Curiosity Notebook, all teaching actions are initiated by the user. Once an action is initiated, the agent will then prompt the user on which information is provided, and in what order. When the teaching action is over, the agent defers to the user by saying, for example, ``You can now select a new button to keep teaching me.''

\begin{figure}[!t]
    \centering
    \includegraphics[width=\linewidth]{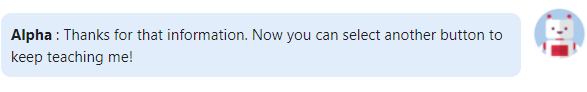}
    \par \smallskip
    \includegraphics[width=\linewidth]{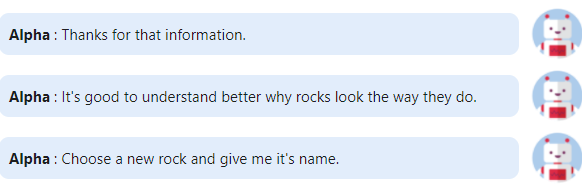}
    \par \smallskip
    \includegraphics[width=\linewidth]{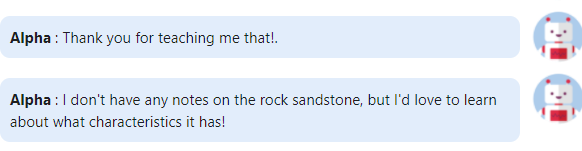}
    \caption{A sample of initiative modes, each begins with the agent thanking the user for their teaching. User initiated action (top), agent initiated action and user rock selection (middle), agent initiated action and rock selection (bottom)}
    \label{initiative_modes}
\end{figure}

The mixed-initiative capabilities were added for both conditions in this experiment, such that when a teaching action was completed, 75\% of the time the user was prompted to select the next teaching action, as in the published version of the notebook, and for the remaining 25\% of the time, the agent chose for themselves, and immediately began the teaching action dialogue. The agent chose between the three teaching actions, describe, explain, and compare, along a probability distribution of [0.5, 0.3, 0.2]. Independently of who initiated,the agent and user each control the decision of which rock to teach, or be taught, 50\% of the time. If fewer than 50\% of all rocks were known to the agent (i.e. something has been taught about them), the agent would choose an unknown rock, otherwise a rock would be chosen at random, which may or may not have been unknown. Examples of the dialogue in these initiation modes are provided in figure \ref{initiative_modes}.

\subsection{Agent Utterances}
\begin{figure}[!t]
    \centering
    \includegraphics[width=\linewidth]{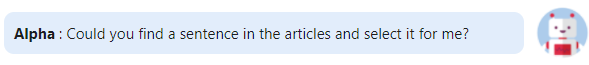}
    \par \smallskip
    \includegraphics[width=\linewidth]{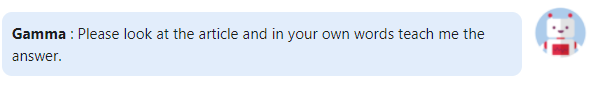}
    \caption{Sample utterances from each agent asking for information from the user, for Alpha (top) and Gamma (bottom)}
    \label{explain_dialog}
\end{figure}

The agent utterances between the two conditions differed when teaching actions were requested. When teaching Alpha, the user would be asked to `select a sentence' to teach, and when teaching Gamma, they would be asked to `read the material and type' their response. An example of this difference is shown in figure \ref{explain_dialog}. Other than this, the utterances were identical between conditions. Each utterance type included several variations in wording, which were chosen at random.

In the event of a learning error, the agent would tell the participant that they didn't understand the input, and they were prompted to confirm whether they would like to try again. After three failed attempts, the agent would automatically move on.

\section{Teaching Modality User Study}
The effect of teaching modality on learning outcomes and engagement was examined by allowing participants to interact with the two different agents, \emph{Alpha} and \emph{Gamma}.

\subsection{Participants}
Participants were recruited from Monash University and personal contacts of the researcher. A total of 46 participants were recruited (23 female), ranging from 20 to 50 years old. Participants were provided with the Explanatory Statement ahead of a mutually agreed-upon time slot. The true nature of the experiment was not disclosed to the participants until the conclusion of the experiment to avoid influencing the participants’ behaviours. This user study was reviewed and received ethics approval through the Monash University Human Research Ethics Committee (Project ID: 26087).

\subsection{Experimental Procedure}
The experiments were conducted as a mix of in-person sessions, and virtual sessions held via Zoom. In both environments, the researcher was present (either on call, or in the room), but did not take part in the teaching interaction, and was only available for technical assistance.

Each participant taught two different agents, Alpha, and Gamma. Participants taught Alpha via sentence selection, and taught Gamma by typed text input. They were told that they would be helping Alpha and Gamma prepare for a test, and were randomly assigned the order in which they would teach the agents. Participants received an introductory explanation of the interface, and of their task, then completed a demographic survey, and a pre-test on rock classification. The test contained two types of questions: those that required users to select all relevant features of a specific rock, and those that asked multiple choice questions about rock formation in general. Participants could spend up to 20 minutes with each agent, and could conduct the teaching interaction as they chose. After interacting with each agent, they completed a post-test on only the six rocks they had just seen, and also a user experience survey. Finally, a survey directly comparing the two agents, and a post-experiment interview was completed.

\subsection{Hypotheses}
The hypotheses of this experiment were:
\begin{itemize}[leftmargin=*]
    \item[] H1: Teaching via text input would result in greater learning gains
    \item[] H2: Teaching via text input would result in higher engagement from the participant
\end{itemize}

\section{User Study Results}

\subsection{Learning Outcomes}
Learning outcomes were measured using the difference between test scores on the pre and post knowledge assessment, per condition. As the pre-test contained questions on all twelve rocks, these scores were split according to the scores achieved for the rocks seen in each condition. The difference in test scores per condition was chosen as a measure of learning outcomes in order to reduce the effects of participant's level of prior knowledge.

Of the 46 participants, five had test results missing. Outliers were detected in the data as those with test scores that fell outside the range of $Q3+1.5\times IQR$ and $Q1-1.5\times IQR$, where Q1 and Q3 are the lower and upper quartile, respectively, and IQR is the inter-quartile range. Two participants were removed under these conditions for this analysis. This gives a total of 44 test differences in the sentence selection condition, and 41 in the text-input condition.

\begin{table}[!t]
\renewcommand{\arraystretch}{1.2}
\caption{Teaching duration metrics per condition}
\label{tab:teaching_attempts}
\resizebox{\linewidth}{!}{%
\begin{tabular}{@{}lll@{}}
\toprule
 & \textbf{Sentence-Selection} & \textbf{Text-Input} \\ \midrule
\textbf{Time per Teaching Action (s)} & $72.5 \pm 19.2$ & $127.0 \pm 54.2$ \\
\textbf{Number of Teaching Attempts} & $19.2 \pm 6.26$ & $12.4 \pm 5.62$ \\ \bottomrule
\end{tabular}%
}
\end{table}

Prior to analysing the differences in test scores, repeated measures ANOVA was conducted to examine the effect of teaching modality on the time taken per teaching action, defined from the time a teaching action was initiated, to the time it was completed (either successfully or unsuccessfully). A significant  difference was found, $F(1,39) = 43.312, p < 0.001$ between the conditions, where participants teaching Gamma took longer, on average, to complete each teaching action, than those teaching Alpha, as seen in Table \ref{tab:teaching_attempts}. This resulted in significantly fewer teaching attempts in the fixed time sessions $F(1,45) = 105.361, p < 0.001$, with Alpha receiving an average of $19.2$ attempts, and Gamma receiving $12.4$ attempts. This is consistent with the requirements of each condition, where teaching via paraphrasing is expected to take longer than teaching via sentence selection, due to the need to both read and type out the paraphrased response.

\begin{table}[!t]
\renewcommand{\arraystretch}{1.2}
\caption{Learning outcomes per condition, defined by the difference between pre and post test results. }
\label{tab:learning_outcomes}
\resizebox{\linewidth}{!}{%
\begin{tabular}{@{}lll@{}}
\toprule
 & \textbf{Sentence-Selection} & \textbf{Text-Input} \\ \midrule
\textbf{Test Score Differences} & $6.39 \pm 4.15$ & $5.90 \pm 2.67$ \\
\textbf{Test Score Differences / Teaching Attempts} & $0.357 \pm 0.272$ & $0.543 \pm 0.354$ \\ \bottomrule
\end{tabular}%
}
\end{table}

Following this, to test H1, repeated measures ANOVA was conducted to examine the effect of teaching modality on the difference in test scores.  

Participants showed a slightly larger difference in test scores in the sentence selection condition compared to the text input condition, as shown in Table \ref{tab:learning_outcomes}, however this difference was not statistically significant, $F(1,38) = 3.371, p = .074$.

Due to the significant difference in the time taken to complete each teaching action, and the fixed duration of the experiment, the test differences were then normalised by the number of teaching attempts, and a repeated measures ANOVA was conducted on these values. For these normalised values, there is a statistically significant difference between conditions, $F(1,38) = 9.039, p = .005$. For tests covering material from the text input condition, participants had a normalised test difference of $0.543$, while for tests covering material from the sentence selection condition, the normalised test difference was $0.357$, shown in Table \ref{tab:learning_outcomes}. The normalised results indicate that learning gains using a text-input modality are higher per teaching action, supporting H1.

\subsection{Engagement}
We next assessed metrics related to participant engagement, firstly those focusing on quantitative metrics of behavioural engagement, and secondly on qualitative measures representing affective engagement.

\subsubsection{Behavioural Engagement}
Behavioural engagement was measured using data representing the participant's interaction with the Curiosity Notebook interface. The metrics used are the number of quizzes initiated by the user, the number of times the agent's notebook was checked, and the number of clicks to navigate between different articles and categories.

As mentioned in the presentation of results on learning outcomes, there was a significant difference in the time taken to complete each teaching action between the two conditions. Participants in the text-input condition tended to spend longer formulating their typed responses, and so had less opportunity to click around the interface in general, or engage in tasks other than teaching, such as quizzes. All metrics have been normalised by the number of teaching attempts to account for this difference. The results are summarised in Table \ref{tab:behaviour_engagement}.

\begin{table}[!t]
\renewcommand{\arraystretch}{1.2}
\caption{Metrics of behavioural engagement, normalised by number of teaching attempts.}
\label{tab:behaviour_engagement}
\resizebox{\linewidth}{!}{%
\begin{tabular}{@{}lll@{}}
\toprule
 & \textbf{Sentence-Selection} & \textbf{Text-Input} \\ \midrule
\textbf{Quizzes / Teaching Attempts} & $0.062 \pm 0.07$ & $0.054 \pm 0.091$ \\
\textbf{Notebook Checks / Teaching Attempts} & $0.272 \pm 0.287$ & $0.373 \pm 0.356$ \\
\textbf{Navigation Clicks / Teaching Attempts} & $1.86 \pm 0.71$ & $1.82 \pm 0.903$ \\ \bottomrule
\end{tabular}%
}
\end{table}

A Wilcoxon signed rank test was used to examine the difference between conditions, as all data was non-normal. This analysis found no statistically significant differences between these normalised measures across the two conditions.

\subsubsection{Affective Engagement}
Metrics of affective engagement were obtained from survey data, administered directly after each teaching interaction, and also after experiencing both conditions. After teaching each agent, a 5-point Likert scale was used to rate enjoyment of the interaction, and perceived utility of the interaction for one's own learning. An IMI survey was used to rate perceived effort and stress during the interaction. The statistically significant results from these surveys are summarised in Table \ref{tab:affective_engagement}.

\begin{table}[!t]
\renewcommand{\arraystretch}{1.2}
\caption{Metrics of affective engagement, per condition}
\label{tab:affective_engagement}
\resizebox{\linewidth}{!}{%
\begin{tabular}{@{}lll@{}}
\toprule
 & \textbf{Sentence-Selection} & \textbf{Text-Input} \\ \midrule
\textbf{Enjoyment} & $3.61 \pm 0.868$ & $3.25 \pm 1.04$ \\
\textbf{Perceived Effort} & $3.49 \pm 0.752$ & $3.71  \pm 0.674$ \\
\textbf{Perceived Utility for Learning} & $3.46 \pm 1.04$ & $3.89  \pm 1.06$ \\ \bottomrule
\end{tabular}%
}
\end{table}

A Wilcoxon signed rank test found a significant difference in ratings of enjoyment between conditions, $p = .0412$, effect size $= 0.277$. Participants rated Alpha slightly higher $(M = 3.61)$ than Gamma $(M = 3.25)$. A significant difference was also found in the participants' perceived effort in teaching task, $p = .02$, effect size $= 0.341$, where Gamma was rated moderately higher $(M = 3.71)$ than Alpha $(M = 3.49)$.

Participants were also asked about their perceived usefulness of the teaching task for their own learning. There was a significant difference between conditions, $p = .0389$, effect size $= 0.279$, where participants rated Gamma slightly higher $(M = 3.89)$ than Alpha $(M = 3.46)$.

There was no statistically significant difference in participant's ratings of their stress $(p = .343)$.

\begin{figure}[!t]
    \centering
    \includegraphics[width=\linewidth]{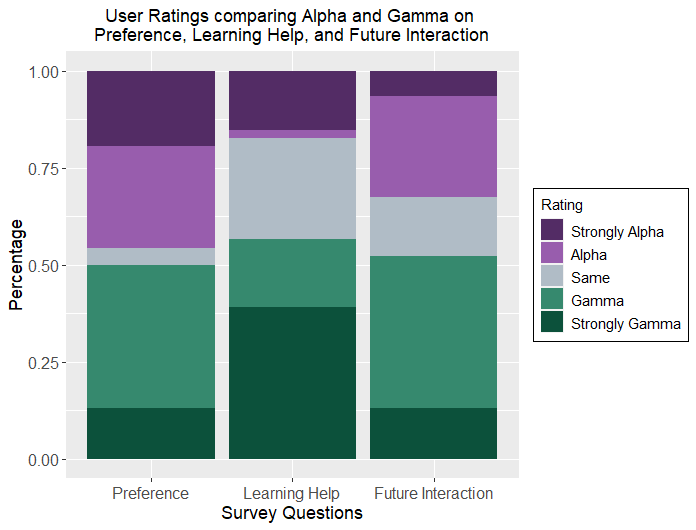}
    \caption{Participant responses to survey questions comparing Alpha and Gamma on preference in teaching, helpfulness for their learning, and which they would like to interact with in future. Results are given as a percentage of participants}
    \label{compare_agents}
\end{figure}

After participants had experienced both conditions, they were given a survey that directly compared their experiences teaching the two agents. They were asked which agent they preferred teaching, teaching which agent was more useful for their own learning, and which agent they would prefer to interact with in the future. A summary of all three survey question results can be seen in figure \ref{compare_agents}.

When asked which agent they preferred teaching, 26.1\% of participants stated that they preferred teaching Alpha, with 19.6\% indicating a strong preference. 37.0\% of participants stated that they preferred teaching Gamma, with 13.0\% indicating a strong preference. The remaining 4.3\% did not have a preference.

When asked which agent they found most helpful for their own learning, 2.2\% of participants felt that Alpha was somewhat more helpful, with 15.2\% indicating that Alpha was much more helpful. 17.4\% of participants felt Gamma was somewhat more helpful, with 39.1\% indicating that Gamma was much more helpful. The remaining 26.1\% found both agents equally helpful.

When asked which agent they would prefer to interact with in the future, 26.1\% of participants indicated they would prefer Alpha, with 6.5\% indicating a strong preference. 39.1\% of participants indicated a preference for Gamma, with 13.0\% indicating a strong preference. The remaining 15.2\% did not have a preference.

\subsubsection{Summary}
Metrics relating to both behavioural engagement and affective engagement have been measured in this study, which provides two different ways through H2 can be evaluated. The results indicate that H2 is supported when we consider affective engagement, but that there is not significant support when we consider behavioural engagement.


\section{Paraphrasing in Text Input Condition}
The analysis of results from the user study indicated that teaching via text input has a positive effect of learning outcomes, and on aspects of affective engagement. During this experimental condition, users were encouraged to paraphrase the source material while formulating their responses. However, there were differences in paraphrasing effort between participants, from fully reformulating the source material, to copying down full sentences from the source material verbatim. Additionally, while learning outcomes in the text input condition were greater than for the sentence selection condition, these results still showed a large variance.  To further examine H1, an analysis was performed on the user inputs to examine the effect of the amount of paraphrasing on learning outcomes, and also on measures of affective engagement.

\subsection{Procedure}
In order to estimate the amount of paraphrasing in each teaching input, semantic similarity between the user generated sentence and the relevant sentence(s) from the source material is used as a proxy. This analysis of user and source data is implemented using Sentence-BERT (SBERT), a network that rapidly derives sentence embeddings that can be compared using cosine similarity \cite{reimers_sentence-bert_2019}. Using this method, if a user typed exactly what was contained in the source sentence, their cosine similarity would be high, indicating a small amount of paraphrasing, and if the cosine similarity is low, this indicates a greater semantic distance between the two sentences, indicating that more paraphrasing has occurred.

Out of all user inputs, only those that returned a matched feature from the Snips natural language model were used. This was done to provide certainty that the user was trying to teach the agent something within scope, and that there would be corresponding material in the source articles with which to compare the user's input. If the user taught about a feature that was recognised by the model, but wasn't relevant to the rock in question, the user input also wasn't processed. The sentence-transformations Python library \cite{reimers_sentence-transformers_nodate} was used to create a model for all usable user inputs, and all source material sentences. Training the model then produced vector sentence encodings for these sentences.

To calculate the cosine similarity between the user input and source material, several situations that may have been present in the user data needed to be accounted for, including teaching:
\begin{itemize}
    \item One feature to the agent, about one rock
    \item Multiple features from a single sentence, about one rock
    \item Multiple features from multiple sentences, about one rock
    \item One feature relevant to multiple rocks
    \item Multiple features relevant to multiple rocks, from one or more sentences per rock
\end{itemize}

In each user teaching input, for the rock it was about, and for each feature that was extracted from the input sentence, all sentences from that rock's source material containing information on that feature were obtained. The cosine similarities between the user input sentence and all relevant source material sentences were calculated. The source sentence with the highest cosine similarity was identified as the most likely sentence the user was paraphrasing from for that feature.

In the cases where multiple features were taught, if the cosine similarity was highest for all features from the same source sentence, only that single sentence was identified. However, if the cosine similarity was highest for different source sentences, those sentences were combined, and the model was retrained with this additional, compound sentence. The cosine similarity was then calculated between the user input sentence and the new, combined sentence.

A similar approach was taken where a user taught information relevant to more than one rock. The sentences with the highest cosine similarity to the user input sentence, per rock, per feature, were identified, and all unique sentences from those identified were combined. The model was retrained on all sentences including the new compound sentences, and the cosine similarity was calculated between the user input sentence and the new compound sentence.

For each user, the average cosine similarity for all their teaching inputs was calculated, and used as an approximation of their paraphrasing effort across the teaching interaction.

\emph{Effect on Learning Outcomes}: Considering the difference in test scores using all questions (`full learning outcomes'), a correlation is present with the cosine similarity, but the relationship is not statistically significant ($r(46) = -0.293, p = .0628$).

\begin{figure}[!t]
    \centering
    \includegraphics[width=\linewidth]{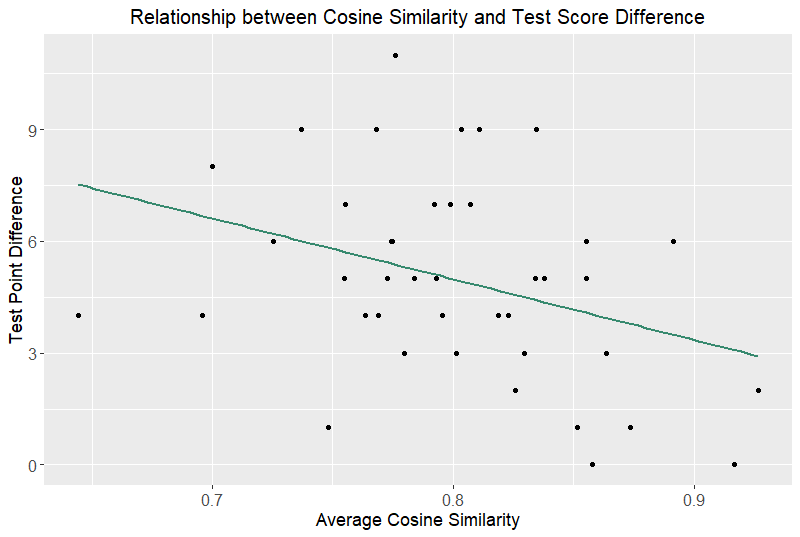}
    \caption{Relationship between average cosine similarity and learning outcomes. The cosine similarity is a proxy for the amount of paraphrasing, where a small value in cosine similarity represents more paraphrasing}
    \label{fig:cos_sim_learning_oucomes}    
\end{figure}

Considering only rock-specific questions ('rock-specific learning outcomes'), a statistically significant correlation is present between cosine similarity and rock-specific learning outcomes ($r(46) = -0.354, p = .0233$). This relationship is shown in Figure \ref{fig:cos_sim_learning_oucomes}.

\emph{Effect on Affective Engagement}:
The Pearson Correlation Coefficient was also calculated for the relationship between the amount of paraphrasing, and the metrics of affective engagement (enjoyment, perceived effort, perceived stress, and perceived usefulness of the interaction for learning).

There were no statistically significant relationships found between the amount of paraphrasing and the metrics of engagement.

\section{Discussion}

\subsection{Learning Outcomes}
This study compares the effect of two different teaching modalities along the metrics of learning outcomes and engagement, during an interaction with a virtual teachable agent. The results indicate that paraphrasing and typing to teach can have a positive effect on both of these metrics. The results also show that there is a significant difference in how long each teaching action takes, and this has an impact on both learning outcomes and engagement. When teaching via paraphrasing and text input, participants were required to read the material, and also take the time to formulate and type out their response. The latter requirement is notably absent when teaching via sentence selection. For both metrics of interest, the increased time taken to complete each teaching action is important due to the fixed duration of the interaction, as fewer teaching actions would be able to occur during the available time.

We see this difference in the results on learning outcomes, where differences in test scores between the two conditions are not statistically significant. However, when we account for the number of teaching actions completed, we see that teaching via paraphrasing and text input resulted in better learning outcomes per teaching action. This suggests that participants were able to better recall the information they taught using this modality. This may be due to a number of reasons, including the increased time spent on on the material during each teaching interaction, the requirement to read the source material more closely, the paraphrasing, the act of typing out the response, or a combination of these. These results also indicate that future experiments using this teaching modality may benefit from removing the time constraint on the teaching interaction time.

The secondary analysis of only the text-input condition to characterise the influence of paraphrasing indicates that the degree to which a participant paraphrased the source material, measured in this case by semantic similarity, has a positive impact on learning outcomes. This adds support to the hypothesis of the study, indicating that not only does paraphrasing and typing the teaching response have a positive impact on learning outcomes, but that more paraphrasing can further these outcomes.

This analysis makes use of the rock-specific learning outcomes, as opposed to the full-learning outcomes. This may show that more paraphrasing is more helpful in questions that focus on knowledge recall. The rock-specific questions ask users to identify features of a rock from a list, while the multi-choice questions ask participants to attribute the explanation of features common across many rocks that they may have seen. It is possible that the broader nature of the multi-choice questions reflected knowledge that the participants already had, and thus did not contribute to a knowledge gain after the teaching interaction took place. 

\subsection{Engagement}

During the study, both behavioural and affective engagement was measured. Metrics of behavioural engagement, representing the participants' interaction with the interface, were not statistically significantly different between the two conditions after normalising for teaching attempts.

In terms of affective engagement, the responses that participants gave differed slightly depending on when they were asked, either immediately following the condition, or at the conclusion of the experiment, when comparing the conditions directly. Responses given directly after the teaching interaction indicate that participants had a preference for teaching Alpha, via sentence selection. Participants also indicated that teaching Gamma, via paraphrasing, required more effort. This suggests that immediately following the interaction, participants were more favourable towards the interaction which required less effort. When participants were asked to directly compare the two conditions, after experiencing them both, there was no clear preference for either condition. This suggests that after some time, any bias that the amount of effort had on reported enjoyment was less prevalent.

Consistently, participants responded that they felt like teaching Gamma was more helpful for their own learning. This supports the hypothesis that participants better learned the material they taught to Gamma, though it is important to note that this study does not explicitly quantify the relationship between perceived versus actual utility for learning.

We also found that a majority of participants would choose to interact with Gamma again, over Alpha. This suggests that there is a priority among this demographic of participants in interacting with the agent that is perceived to have a greater impact on learning, rather than the one that was more enjoyable, or less effort, to teach.

The secondary analysis focusing on the amount of paraphrasing found no significant impact on measures of affective engagement. This may be interpreted positively, as it indicates that paraphrasing more extensively from the source material was not perceived as requiring more or less effort, as being more or less stressful, and either increased nor decreased enjoyment of the teaching task, compared to more limited, or no paraphrasing. The differences in measures of affective engagement when compared to the sentence selection condition may only be a reflection of the time and effort required to type out the teaching response at all, thus encouraging users to paraphrase more may benefit their learning outcomes at limited detriment to their user experience.

It is also noteworthy that despite this analysis indicating that the amount of paraphrasing did positively impact learning outcomes, participants did not perceive this effect themselves. This may be a reflection on the design of the knowledge assessments, where participants were not able to accurately assess their performance, or draw meaningful connections between the material they taught the agent, and the material they were tested on.

\subsection{Limitations of Study and Analysis}

There are some limitations in the study design that should be considered when contextualising the results. This study examined learning outcomes, but only focused on short-term recall, and did not investigate the effects of teaching modality on long term knowledge retention, which may be of interest if the technology were to be deployed in a classroom setting.

The fixed duration of the experiment also poses some limitations, including those already discussed in relation to the pace of teaching in each condition. Additionally, constraining the amount of time participants can spend in each condition may have placed pressure on participants to maximally cover the material in the available time, and impacting teaching choices, perhaps to favour wider coverage over depth.

Finally, using semantic similarity as a proxy for the amount of paraphrasing of the teaching utterances, compared to the source material places limitations on the conclusions drawn from this analysis. Most significantly, this measure does not include any indication of the quality of the paraphrase, or of the accuracy of the information from a teaching perspective. For example, the input sentence may be semantically distant from the source material because it is brief, lacking in detail, or contains information outside the source article.


\section{Conclusion and Future Work}
In this work, we investigate the effect of teaching modality while interacting with a virtual teachable agent, via an online teaching platform, the Curiosity Notebook. A method of selecting sentences from the source material is compared against a method of reading and paraphrasing the source material, and typing out responses to teach. A user study has been conducted to measure the learning outcomes and engagement of participants across the two conditions. A secondary analysis conducted on the paraphrasing condition alone to measure the effect of differences in individual paraphrasing behaviours.


The results of the user study indicate that teaching via paraphrasing does have a positive effect on learning outcome, improving recall of the material covered. It was observed that individual teaching actions using this modality took longer to complete than for teaching via sentence selection, which  reduced the total volume of material covered when teaching via paraphrasing.

This also affected metrics of behavioural engagement, as users experienced periods of inactivity while reading or typing. Teaching via paraphrasing was recognised as requiring more effort, however this did not significantly negatively impact enjoyment of the teaching task. Additionally, it was consistently recognised that teaching via paraphrasing was more helpful for the users' own learning, which positively affected their desire to use this teaching modality in the future.

It was also found that the more paraphrasing in the participants' teaching inputs, measured via semantic similarity, had a positive impact on learning outcomes. Conversely, the amount of paraphrasing did not have any effect on the perceived effort or enjoyment of the teaching task, indicating that this benefit to learning outcomes does not come at the expense of these affective measures.

This teaching interaction involves other factors, such as the error rate in responses of the agent, and the teaching path selected by individual users. Future development of the Curiosity Notebook that utilises teaching via paraphrasing would benefit from exploration of the factors to determine which have an effect on the metrics of interest, and how they may be optimised to benefit students.

\bibliographystyle{IEEEtran}
\bibliography{IEEEabrv,references}

\end{document}